\documentclass[prl,twocolumn]{revtex4}
\pdfoutput=1
\usepackage{textcomp}
\usepackage{amssymb}
\usepackage{anysize}
\usepackage{bold-extra}
\usepackage{enumerate}
\usepackage{graphicx}
\usepackage{lscape}
\usepackage{mathrsfs}
\usepackage{multirow}
\usepackage[final]{pdfpages}
\usepackage{rotating}
\usepackage{subfig}
\usepackage{url}
\usepackage{wrapfig}
\usepackage{amsmath}
\usepackage{color}
\usepackage{fancyhdr}
\usepackage{verbatim} 
\usepackage[linktocpage,colorlinks]{hyperref}

\def\bea{\begin{eqnarray}}
\def\eea{\end{eqnarray}}
\def\ba{\begin{eqnarray}}
\def\ea{\end{eqnarray}}
\def\be{\begin{equation}}
\def\ee{\end{equation}}
\def\beq{\begin{equation}}
\def\eeq{\end{equation}}

\def\lsim{\mbox{\raisebox{-.6ex}{~$\stackrel{<}{\sim}$~}}}
\def\gsim{\mbox{\raisebox{-.6ex}{~$\stackrel{>}{\sim}$~}}}

\newcommand{\slashed}{\slash \hspace{-0.23cm}}

\begin{document}

\preprint{UCSD/PTH 13-01}
\title{Bottom-Quark Forward-Backward Asymmetry in the Standard Model and Beyond}
\author{Benjam\'{i}n Grinstein}
\email{bgrinstein@ucsd.edu}
\affiliation{Department of Physics, University of California, San Diego, La Jolla, CA 92093 USA}
\author{Christopher W. Murphy}
\email{cmurphy@physics.ucsd.edu}
\affiliation{Department of Physics, University of California, San Diego, La Jolla, CA 92093 USA}

\begin{abstract}
We computed the bottom-quark forward-backward asymmetry at the Tevatron in the Standard Model and for several new physics scenarios.  Near the $Z$-pole, the SM bottom asymmetry is dominated by tree level exchanges of electroweak gauge bosons.  While above the $Z$-pole, next-to-leading order QCD dominates the SM asymmetry as was the case with the top quark forward-backward asymmetry.  Light new physics, $M_{NP} \lsim 150$ GeV, can cause significant deviations from the SM prediction for the bottom asymmetry. The bottom asymmetry can be used to distinguish between competing NP
explanations of the top asymmetry based on how the NP interferes
with $s$-channel gluon and $Z$ exchange.
\end{abstract}

\maketitle


\section{Introduction}
Measurements~\cite{Aaltonen:2011kc, Abazov:2011rq, Aaltonen:2012it} of the forward-backward asymmetry in top-quark pair production ($A^{t\bar{t}}_{FB}$) by the CDF and D\O\, collaborations at the Tevatron have attracted a lot of attention recently.  At high invariant mass, the CDF measurement $A^{t\bar{t}}_{FB}(M_{t\bar{t}} \geq 450\, \text{GeV}) = 0.295 \pm 0.058 (\text{stat.}) \pm 0.031 (\text{syst.})$ is approximately 3$\sigma$ away from the Standard Model (SM) prediction, $0.100 \pm 0.030$~\cite{Aaltonen:2012it}.  In addition, CDF observes that $A^{t\bar{t}}_{FB}$ has an approximately linear dependence on both the invariant mass and the magnitude of the rapidity difference ($|\Delta y_{t\bar{t}}|$) of the $t\bar{t}$ pair with slopes that are more than 2$\sigma$ away from the SM prediction.  

Soon after CDF reported evidence for a mass-dependent $t\bar{t}$ asymmetry, it was realized~\cite{Bai:2011ed, Strassler:2011vr, Kahawala:2011sm} that measuring the forward-backward asymmetry in bottom quark production ($A^{b\bar{b}}_{FB}$) may provide insight into the source of the $t\bar{t}$ asymmetry.  Any new physics (NP) explanation of $A^{t\bar{t}}_{FB}$ involving left- (right-)handed quarks that respects $SU(2)_L$ (custodial) symmetry will in general also create an asymmetry in $b\bar{b}$ production.  The CDF collaboration is in the process of measuring the $b\bar{b}$ forward-backward asymmetry, and has stated~\cite{website:bartos} how it is binning the data and how sensitive it expects to be to a potential signal.  However, $A^{b\bar{b}}_{FB}$ will likely be more difficult to measure than $A^{t\bar{t}}_{FB}$.  Among the reasons for this are that gluon fusion, which does not produce an asymmetry, is responsible for $\gsim 90\%$ of bottom quark production at the Tevatron.  In addition, the $b\bar{b}$ asymmetry is measured by selecting dijet events containing a soft muon, and relating the charge of the muon to the charge of the $b$ that produced it~\cite{website:bartos}.  This is potentially problematic because $B - \bar{B}$ mixing and cascade decays will partially wash out the correlation between the charge of what is detected and the charge of the bottom quark that produced it~\cite{Sehgal:1987wi}.

In this Letter, we computed the bottom-quark forward-backward asymmetry at the Tevatron in the SM and for several NP scenarios.  It is necessary to know the SM prediction in order to determine whether or not any NP can possibly be present.  Since a small asymmetry is expected in the SM, $A_{FB}$ provides an excellent window to observe NP.  An interesting difference between the bottom and top quark asymmetries is that the $Z$-pole is in the signal region for the $b\bar{b}$ asymmetry. This leads to tree level exchanges of electroweak gauge bosons dominating the SM contribution to $A_{FB}$ near the $Z$-pole, as well as the opportunity for there to be significant interference effects between NP and tree level $Z$ exchange. 

\section{Standard Model Calculation}
\label{sec:sm}
%
The definition of the forward-backward asymmetry in heavy quark production we use is
\begin{equation}
A_{FB} = \frac{\sigma(\Delta y > 0) - \sigma(\Delta y < 0)}{\sigma(\Delta y > 0) + \sigma(\Delta y < 0)}.
\end{equation}
Here $\Delta y$ is the difference in the rapidity of the quark and anti-quark, $\Delta y \equiv y_Q - y_{\bar{Q}}$, and is invariant under boosts along the collision axis.  A frame dependent asymmetry may also be defined using $y_Q$ instead of $\Delta y$ as the discriminating observable.  Leading order (LO) QCD is completely symmetric with respect to $\Delta y$, and thus does not generate an asymmetry.  Starting with next-to-leading order (NLO) QCD, contributions to the asymmetry as an expansion in powers of $\alpha_s$ can be written schematically as
\begin{align} \label{eq:asymexp}
A_{FB} &= \frac{N}{D} = \frac{\alpha^2 \tilde{N}_0 + \alpha_s^3 N_1 + \alpha_s^2 \alpha \tilde{N}_1 + \alpha_s^4 N_2 + \cdots}{\alpha_s^2 D_0 + \alpha^2 \tilde{D}_0 + \alpha_s^3 D_1 + \alpha_s^2 \alpha \tilde{D}_1 + \cdots} \nonumber \\
&= \alpha_s \frac{N_1}{D_0} + \frac{\alpha^2}{\alpha_s^2}\frac{\tilde{N}_0}{D_0} + \alpha\frac{\tilde{N}_1}{D_0}  + \cdots .
\end{align}
Analytic formulae for the $\mathcal{O}(\alpha_s)$ and $\mathcal{O}(\alpha)$ terms of $A_{FB}$ are given in~\cite{Kuhn:1998jr, Kuhn:1998kw}.  These results are based on analogous calculations~\cite{Berends:1973fd, Berends:1982dy} for the $e^- e^+ \rightarrow \gamma^{\star} \rightarrow \mu^- \mu^+$ asymmetry.  Prior results on the QCD asymmetry also exist~\cite{Brown:1979dd, Ellis:1986ef, Halzen:1987xd}.  The $\mathcal{O}(\alpha^2 / \alpha_s^2)$  term for $A^{t\bar{t}}_{FB}$ was computed in~\cite{Hollik:2011ps}.  Electroweak (EW) Sudakov corrections are shown in~\cite{Manohar:2012rs} to increase the $\mathcal{O}(\alpha_s)$ contribution to the inclusive $A^{b\bar{b}}_{FB}$ by a factor of 1.07.  While the $N_1$ and $D_1$ terms in~\eqref{eq:asymexp} are known completely and have been studied~\cite{Almeida:2008ug, Dittmaier:2008uj, Melnikov:2009dn, Kidonakis:2011zn, Kuhn:2011ri, Alioli:2011as, Melnikov:2011qx, Campbell:2012uf, Skands:2012mm, Bernreuther:2012sx} in depth, $N_2$ is only partially known~\cite{Ahrens:2011mw, Ahrens:2011uf}.\footnote{See~\cite{Ahrens:2010zv, Baernreuther:2012ws, Gao:2012ja, Brucherseifer:2013iv} for some beyond NLO calculations of symmetric heavy quark observables.} Since it would be inconsistent to include the $N_1 D_1 / D_0$ term in our calculation without the $N_2$ term, we drop the $\mathcal{O}(\alpha_s^2)$ contribution to $A_{FB}$.  To account for this neglect of higher order terms, we assign an uncertainty to our calculation of 30\% of the $\mathcal{O}(\alpha_s)$ contribution, originating from $\alpha_s D_1 \approx 0.3 D_0$.

Our calculation was done by convolving the analytic formulae of~\cite{Kuhn:1998kw, Hollik:2011ps} with MSTW 2008 NLO PDFs~\cite{Martin:2009iq} using the deterministic numeric integration algorithm Cuhre from the CUBA library~\cite{Hahn:2004fe}.  $\alpha_s$ is set by the MSTW2008 best-fit value, $\alpha_s (M_Z) = 0.120$.  We fixed $\mu_R = \mu_F = M_Z$ and $n_{lf} = 4$.  The other numeric values employed in this analysis were: $m_b = 4.7$ GeV, $M_Z = 91.1876$ GeV, $\Gamma_Z = 2.4952$ GeV, $\alpha(M_Z) = 1/128.93$, and $\sin^2\theta_W = 0.231$.

To mimic CDF's analysis~\cite{website:bartos} we required the
$b\bar{b}$ pair in our calculation to have a maximum acollinearity of
$\delta = \pi - 2.8$ radians.  The phase space that is available to the gluon in the $b\bar
b g$ final state is discussed in~\cite{Berends:1973tz}.   Additional cuts, $|y_{b,\bar{b}}| \leq 1$, and  $p_{\perp b,\bar{b}} \geq 15$ GeV were made.
We found the $\mathcal{O}(\alpha)$ corrections decrease the
contribution of $\mathcal{O}(\alpha_s)$ to $A^{b\bar{b}}_{FB}$ by
3-11\%, depending on the bin.  However, we neglect this
$\mathcal{O}(\alpha)$ contribution as it is mostly canceled by
the increase in $A^{b\bar{b}}_{FB}$ due to electroweak Sudakov
effects~\cite{Manohar:2012rs}, and the sum of the two effects is small
compared to the uncertainty in the total contribution.  The flavor
excitation process, $q g \rightarrow q b \bar{b}$, as well as
$t$-channel $W$ exchange were also neglected as they are numerically
small~\cite{Kuhn:1998kw, Hollik:2011ps}.

Our results for the $\mathcal{O}(\alpha^2 / \alpha_s^2)$ and
$\mathcal{O}(\alpha_s)$ contributions to binned $A^{b\bar{b}}_{FB}$
are shown in Table~\ref{tab:bbbarasym}.  In the second and third
columns the uncertainty is due to varying $\mu_R=\mu_F$ from $M_Z / 2$ to
$2 M_Z$.  In the fourth column the first uncertainty is due to neglect
of higher-order terms, and the second is the combined scale
uncertainty.  The uncertainty in the $\mathcal{O}(\alpha^2 /
\alpha_s^2)$ contribution to $A^{b\bar{b}}_{FB}$ is larger than the
$\mathcal{O}(\alpha_s)$ term because the extra power of $\alpha_s$
makes it more sensitive to the choice of scales and PDFs.
\begin{table*}
\centering
 \begin{tabular}{  c  c c  c }
 Bin & $\mathcal{O}(\alpha^2 / \alpha_s^2)$ & $\mathcal{O}(\alpha_s)$  & $A^{b\bar{b}}_{FB}[\%]$  \\ \hline\hline
  $35 \leq M_{b\bar{b}}/\text{GeV} < 75$ & 0.                                             & $0.179^{+0.014}_{-0.011}$ & $0.18 \pm 0.05\,^{+0.01}_{-0.01}$ \\
  $75 \leq M_{b\bar{b}}/\text{GeV} < 95$ & $2.167^{+0.661}_{-0.550}$ & $0.676^{+0.032}_{-0.026}$ &  $2.84 \pm 0.20\,^{+0.69}_{-0.58}$ \\
$95 \leq M_{b\bar{b}}/\text{GeV} < 130$ & $0.554^{+0.178}_{-0.147}$ & $1.241^{+0.058}_{-0.048}$ &  $1.79 \pm 0.37\,^{+0.24}_{-0.20}$ \\
        $130 \leq M_{b\bar{b}}/\text{GeV} $ & $0.150^{+0.046}_{-0.039}$ & $3.369^{+0.237}_{-0.199}$  & $3.52 \pm 1.01\,^{+0.28}_{-0.24}$ \\ \hline
      $0.0 \leq |\Delta y_{b\bar{b}}| < 0.5$ & $0.023^{+0.005}_{-0.005}$ & $0.032^{+0.002}_{-0.001}$  & $0.06 \pm 0.01\,^{+0.01}_{-0.01}$ \\
      $0.5 \leq |\Delta y_{b\bar{b}}| < 1.0$ & $0.082^{+0.020}_{-0.017}$ & $0.166^{+0.012}_{-0.010}$ & $0.25 \pm 0.05\,^{+0.03}_{-0.03}$ \\
   $1.0 \leq |\Delta y_{b\bar{b}}| \leq 2.0$ & $0.133^{+0.034}_{-0.029}$ & $0.382^{+0.031}_{-0.024}$  & $0.51 \pm 0.11\,^{+0.07}_{-0.05}$ \\ \hline
                                                      Inclusive & $0.074^{+0.018}_{-0.015}$ & $0.226^{+0.021}_{-0.016}$ &  $0.30 \pm 0.07\,^{+0.04}_{-0.03}$ \\ \hline\hline
  \end{tabular}
  \caption{The $\mathcal{O}(\alpha^2 / \alpha_s^2)$ and $\mathcal{O}(\alpha_s)$ contributions to $A^{b\bar{b}}_{FB}$ in various bins.}
  \label{tab:bbbarasym}
\end{table*}

Based on CDF's expected sensitivities~\cite{website:bartos} and
assuming the Standard Model (and the measurements follow a Gaussian
distribution), CDF should be able to exclude $A^{b\bar{b}}_{FB}(75 \leq
M_{b\bar{b}}/\text{GeV} < 95) = 0$ at the 2.2$\sigma$ confidence level
(CL).  Although the central value for the asymmetry in the $\geq130$
GeV invariant mass bin is slightly larger than the $75-95$ GeV bin,
CDF should only be able to exclude $A^{b\bar{b}}_{FB}(130 \leq
M_{b\bar{b}}/\text{GeV}) = 0$ at the 1.2$\sigma$ CL.  The likelihood
of excluding zero asymmetry in the $95-130$ GeV invariant mass bin is
comparable to the likelihood in the $\geq130$ GeV bin.  In the SM, all
the other (mass or rapidity) bins should be consistent with zero at
the 1$\sigma$ level based on experimental uncertainties.

LO event generators can predict the $\mathcal{O}(\alpha^2 / \alpha_s^2)$ contribution to the asymmetry.  MadGraph 5.1.5.5~\cite{Alwall:2011uj} with CTEQ6L1 PDFs~\cite{Pumplin:2002vw} gives $A^{b\bar{b}}_{FB}(75 \leq M_{b\bar{b}}/\text{GeV} < 95) = (2.26 \pm 0.32(\text{stat.})^{+0.24}_{-0.74}(\text{scale}))\%$, in good agreement with our calculation.

It has been suggested~\cite{ Strassler:2011vr, Kahawala:2011sm} that measuring the charm-quark forward-backward asymmetry at the Tevatron ($A^{c\bar{c}}_{FB}$) and the bottom-quark charge asymmetry at the LHC ($A^{b\bar{b}}_{C}$) may also provide insight into the origin of the $A^{t\bar{t}}_{FB}$ anomaly.  We computed SM asymmetries of a few percent in suitably chosen kinematic regions for both $A^{c\bar{c}}_{FB}$ and $A^{b\bar{b}}_{C}$.  While the central values for these asymmetries are comparable to those of $A^{b\bar{b}}_{FB}$, it is unlikely that these asymmetries will be observed any time soon in the absence of NP.  For $A^{c\bar{c}}_{FB}$, $c$-tagging is less efficient than $b$-tagging.  For $A^{b\bar{b}}_{C}$, the kinematic regions where the asymmetry becomes a few percent have small production cross sections, and will require the LHC to run for at least a year at 14 TeV to collect enough data for the SM asymmetry to be statistically distinguishable from zero.  Furthermore, the EW contribution to the cross section in these kinematic regions is negligible, and no $Z$-resonance effects are expected.

\def\OMIT#1{{}}
\section{New Physics Scenarios}
\label{sec:np}
%
Many new physics models have been proposed~\cite{Stone:2011dn,
  Grinstein:2012pn, Tavares:2011zg, AguilarSaavedra:2011ci, Krnjaic:2011ub, Blum:2011fa,
  Grinstein:2011yv, Grinstein:2011dz, Alvarez:2010js, Djouadi:2011aj,
  Delaunay:2012kf} as explanations of the anomalously\footnote{For a
  contrary view, see~\cite{Brodsky:2012ik} where it is argued
  this discrepancy is not a signal of NP, but is instead due to
  uncertainty in the choice of which renormalization scale should be
  used.} large $t\bar{t}$ forward-backward asymmetry.  For the
stringent constraints that these models must overcome
see~\cite{Drobnak:2012cz, Gross:2012bz, Gresham:2012kv, Cvetic:2012kv,
  Gresham:2012wc, Alvarez:2012ca, Drobnak:2012rb, Knapen:2011hu, AguilarSaavedra:2011hz, AguilarSaavedra:2011ug}.  Prospects for discovery at the LHC are discussed in~\cite{Grinstein:2012pn, Tavares:2011zg, Grinstein:2011dz, Djouadi:2011aj,  Delaunay:2012kf, Drobnak:2012cz, Gross:2012bz, Gresham:2012kv, Alvarez:2012ca, Drobnak:2012rb, Knapen:2011hu, AguilarSaavedra:2011hz, AguilarSaavedra:2011ug} among others.
\OMIT{e predict the NP contribution to the parton level $b\bar{b}$
  forward-backward asymmetry at the Tevatron, and update the
  predictions for $A^{t\bar{t}}_{FB}$ in light of the release of the
  full dataset analysis.  While some} Predictions for
$A^{b\bar{b}}_{FB}$ in the context of various NP scenarios have
already been made in~\cite{Sehgal:1987wi, Saha:2011wr, Kahawala:2011sm, Drobnak:2012cz,
  Delaunay:2012kf, Ipek:2013zi}.  We expanded on these works by taking
into account the resonance effects of the $Z$, and limiting
ourselves to the energy regime accessible at the Tevatron.  In particular, we are interested in seeing if the NP
contribution to $A^{b\bar{b}}_{FB}$ can be large enough to be distinguishable from the SM predictions we computed above
based on the expected sensitivities given in~\cite{website:bartos}.  Any NP in
the bottom sector must not spoil the agreement between the SM and
precise measurements of flavor changing decays and meson mixing
observables such as Br$(b \rightarrow s + \gamma)$ and $B - \bar{B}$
mixing.  These and other constraints, such as same-sign top
production, are more easily satisfied in flavor symmetric models in
which the NP particles form complete representations of the quark
global flavor symmetry group, $G_F = SU(3)_{U_R}\times
SU(3)_{D_R}\times SU(3)_{Q_L}$.  Furthermore, the flavor symmetry
guarantees a definite relationship between $A^{t\bar{t}}_{FB}$ and
$A^{b\bar{b}}_{FB}$.  We consider three different models, a light, broad
axigluon $(G^{\prime})$, a scalar weak doublet $(\phi)$, and an
$SU(3)_{Q_L}$ octet of electroweak triplet (EWT) vectors ($V$); see Table~\ref{tab:symreps}.
\begin{table*}
\centering
 \begin{tabular}{  c  c c c c }
  Case & SM & $ G_F$ & Relevant Interaction & Ref. \\ \hline \hline
$G^{\prime}$ & $(8,1)_0$ & (1,1,1) & $g_a \left(\bar{U}_{R}\, \slashed{G}^{\prime}\, U_{R} + \bar{D}_{R}\, \slashed{G}^{\prime}\, D_{R}  - \bar{Q}_{L}\, \slashed{G}^{\prime}\, Q_{L}\right)$  & \cite{Tavares:2011zg, Krnjaic:2011ub} \\
$\phi$ & $(1,2)_{1/2}$ & $(3,1,\bar{3})$ & $\lambda \left(\phi^0\, \bar{t}_{L}\, V_{tb}\, u_{R} + \phi^-\, \bar{b}_{L}\, u_{R}\right)$ + h.c. & \cite{Blum:2011fa}\\
$V$ & $(1,3)_0$ & (1,1,8) & $\eta\, V^{a,b}_{\mu} \left(\bar{Q}^{\alpha i}_{L}\, \gamma^{\mu}\, (T^a_{Q})^{\beta}_{\alpha}\, (T^b_L)^j_{i}\, Q_{L\beta j}\right)$ & \cite{Grinstein:2011yv, Grinstein:2011dz} \\ \hline\hline
  \end{tabular}
  \caption{The gauge and flavor representations for the models under consideration. $T^a_{Q}$ and $T^b_{L}$ are generators of $SU(3)_{Q_L}$ and $SU(2)_L$ respectively.}
  \label{tab:symreps}
\end{table*}

It is convenient to split the contributions to the forward-backward asymmetry into two terms
\begin{equation} \label{eq:asymNP}
A_{FB} = A^{\text{I}}_{FB} + A^{\text{II}}_{FB}.
\end{equation}
$A^{\text{I}}_{FB}$ contains the $\mathcal{O}(\alpha_s)$ contribution
to $A^{b\bar{b}}_{FB}$, and can be obtained from
Table~\ref{tab:bbbarasym}.  The $\mathcal{O}(\alpha)$ contribution to
the asymmetry could also be included in $A^{\text{I}}_{FB}$, but we
neglect it in what follows. On the other hand, $A^{\text{II}}_{FB}$
contains the SM $\mathcal{O}(\alpha^2 / \alpha_s^2)$ contribution to
the asymmetry as well as contributions from NP.  This includes both
pure NP contributions and interference between NP and tree level
$s$-channel gluon and $Z$ exchange.  We calculated $A^{\text{II}}_{FB}$
using FeynRules 2.0.24~\cite{Alloul:2013bka} to implement the NP
models in MadGraph 5.1.5.5~\cite{Alwall:2011uj} including electroweak
processes (\texttt{QED=2}).  
For $A^{t\bar{t}}_{FB}$, $10^5$ events were generated for a given set of
parameters using the CTEQ6L1~\cite{Pumplin:2002vw} PDFs with the
renormalization and factorization scales set to $m_t$.  For $A^{b\bar{b}}_{FB}$, $10^5$ events were generated for each mass bin for a
given set of parameters with $\mu_R = \mu_F = M_Z$.  As was the case
for the SM analysis, a cut was placed on the rapidity of the bottom
quarks, $|y_{b,\bar{b}}| \leq 1$.

\begin{figure*}
  \centering
 \subfloat{\label{fig:axi_tt}\includegraphics[width=0.5\textwidth]{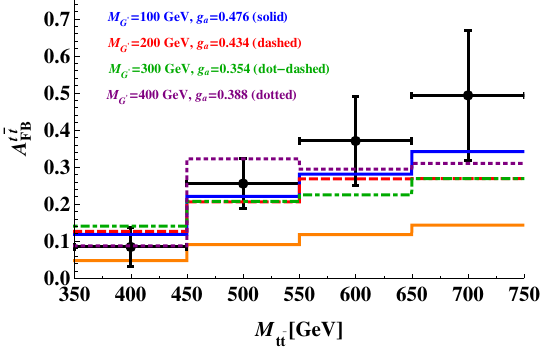}} 
  \subfloat{\label{fig:axi_bb}\includegraphics[width=0.5\textwidth]{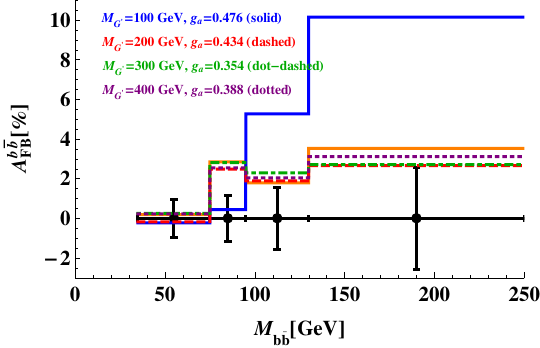}} \\
 \subfloat{\label{fig:light_tt}\includegraphics[width=0.5\textwidth]{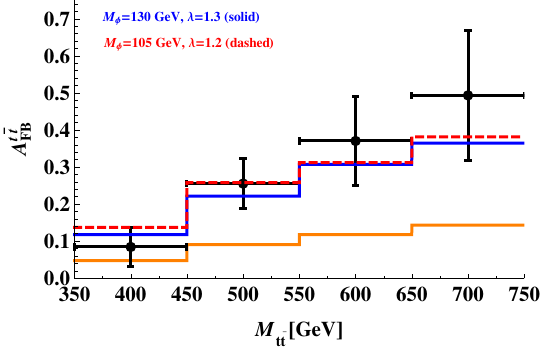}} 
  \subfloat{\label{fig:light_bb}\includegraphics[width=0.5\textwidth]{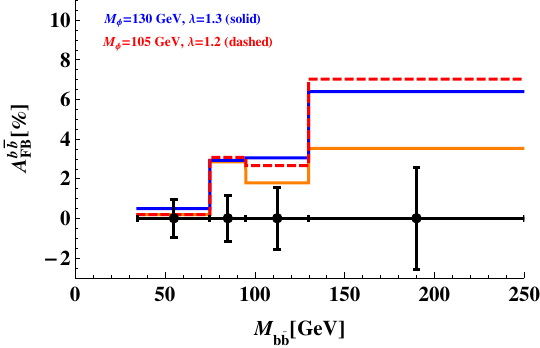}} \\
 \subfloat{\label{fig:vec_tt}\includegraphics[width=0.5\textwidth]{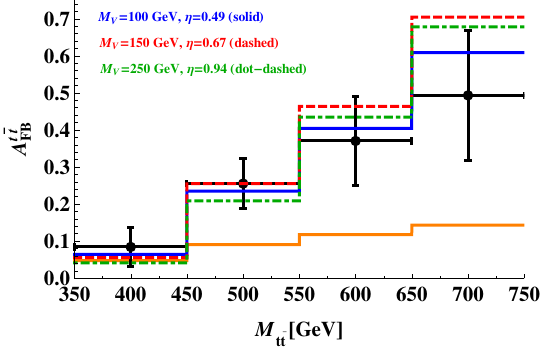}} 
  \subfloat{\label{fig:vec_bb}\includegraphics[width=0.5\textwidth]{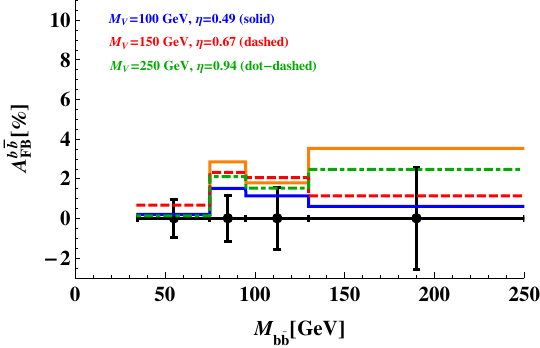}} \\
   \caption{Predictions for the binned $A^{t\bar{t}}_{FB}$ (left) and $A^{b\bar{b}}_{FB}$ (right) from the axigluon (top), scalar weak doublet (middle), and  flavor octet vector (bottom) models.  SM predictions are in orange.  In black are CDF's measurements~\cite{Aaltonen:2012it} and expected sensitivities~\cite{website:bartos} for $A^{t\bar{t}}_{FB}$ and $A^{b\bar{b}}_{FB}$ respectively.}
 \label{fig:asym}
   \end{figure*}

Predictions for the binned $t\bar{t}$ and $b\bar{b}$ asymmetries from
the NP models are shown in the left and right columns of
Figure~\ref{fig:asym} respectively.  
\OMIT{The top, middle, and bottom rows
are the axigluon, weak scalar doublet, and flavor symmetric vector
respectively.  Shown in orange are the SM predictions for the
$t\bar{t}$ and $b\bar{b}$ asymmetries from Ref.~\cite{Aaltonen:2012it}
and Sec.~\ref{sec:sm} respectively.  Shown in black are CDF's
measurements including error bars~\cite{Aaltonen:2012it} and expected
sensitivities~\cite{website:bartos} for $A^{t\bar{t}}_{FB}$ and
$A^{b\bar{b}}_{FB}$ respectively.} Overflow is included in the
rightmost bins. The widths of the axigluon and the EWT vectors were chosen
to be 10\% of their masses.  For the scalars, the natural width to
quarks was used.  Axigluon benchmark points were taken from Table I
of~\cite{Gross:2012bz}.   Benchmark points for the $\phi$ and $V$
models were chosen based on adding approximately 10\% to the inclusive
$t\bar{t}$ asymmetry, having a roughly linear dependence of
$A^{t\bar{t}}_{FB}$ on $M_{t\bar{t}}$, and adding (or subtracting)
less than 1~pb from the $t\bar{t}$ production cross section at the
Tevatron.

We have given three classes of models that can accommodate $A^{t\bar{t}}_{FB}$ and produce a $A^{b\bar{b}}_{FB}$ that is distinguishable from the SM prediction.  However, this is not generally the case.  For example, a flavor octet, EW singlet model ($V^b_{\mu}\,(T^b_L)^j_i \rightarrow V_{\mu}\delta^j_i$ in Table~\ref{tab:symreps}), can accommodate $A^{t\bar{t}}_{FB}$ without causing any significant deviations from the SM predictions because it only produces $b\bar{b}$ from $d\bar{d}$ initial states whereas the other models involve $u\bar{u}$ initial states.  While all three models considered can interfere with gluon exchange, $\phi$ and $V$ can also interfere with the $Z$, which dominates the NP contribution to $A^{b\bar{b}}_{FB}$ in the $Z$-pole bin for these models.

\OMIT{To generate an asymmetry distinguishable from the SM, the NP needs to couple $b\bar{b}$ to $u\bar{u}$, only coupling $b\bar{b}$ to $d\bar{d}$ will not suffice. This can occur through neutral $s$-channel exchange ($G^{\prime}$), charged $u$-channel exchange ($\phi$), or a combination of the two ($V$).  In contrast, a flavor octet, EW singlet model ($V^b_{\mu}\,(T^b_L)^j_i \rightarrow V_{\mu}\delta^j_i$), which can only contribute to $A^{b\bar{b}}_{FB}$ through neutral $t$-channel exchange, does not cause any significant deviations from the SM predictions.  While all three models considered can interfere with gluon exchange, $\phi$ and $V$ can also interfere with the $Z$, which dominates the NP contribution to $A^{b\bar{b}}_{FB}$ in the $Z$-pole bin for these models.
The axigluons contributes positively (negatively) to the asymmetry for $M_{b\bar{b}} >(<)~M_{G^{\prime}}$ due to interference with gluon exchange.  A negative 2\% asymmetry in the $Z$-pole bin due to interference with the $Z$ is the most striking prediction from the scalar weak doublet.    On the other hand, the neutral vector contributions to the asymmetry, including interference with $Z$ and gluon exchange, mostly cancel and do not cause any significant deviations from the SM prediction for $A^{b\bar{b}}_{FB}$.}

In addition to the $A^{t\bar{t}}_{FB}$ anomaly, there is the longstanding puzzle of the $b\bar{b}$ forward-backward asymmetry at LEP1, $A^{(0,b)}_{FB}$, which is 2.4$\sigma$ below the SM value~\cite{Beringer:2012}.  Furthermore, the ratio of the partial width $Z \rightarrow b\bar{b}$ to the inclusive hadronic width, $R_b$, is 2.3$\sigma$ above the SM prediction~\cite{Freitas:2012sy}.  Assuming only the bottom quark's coupling to the $Z$ is modified, the value of $\delta g_{Rb}$ which provides the best-fit to the EWPD collected at LEP is 0.016~\cite{Batell:2012ca}, which is more than 20\% of the LO SM coupling.  See~\cite{Alvarez:2010js, Djouadi:2011aj} for attempts to simultaneously explain $A^{t\bar{t}}_{FB}$  and $A^{(0,b)}_{FB}$.  In models where the NP couples to quarks in a flavor universal way, the loop correction that gives the best-fit value for $\delta g_{Rb}$ will give an analogous correction to $\delta g_{Ru,d}$, which is much larger than allowed by atomic parity violation experiments~\cite{Gresham:2012wc}.  The tree level $V-Z$ mixing of~\cite{Grinstein:2011gq} is not a viable explanation either for the same reason.    Axigluon models give $\delta g_{Rb} = \delta g_{Lb}$~\cite{Gresham:2012wc}, which disagrees with the best-fit value for $\delta g_{Lb}$, $\mathcal{O}(10^{-3})$~\cite{Batell:2012ca}.  Prospects for measuring $b\bar{b}$ and $t\bar{t}$ asymmetries at future linear colliders are examined in~\cite{Guo:2013dc}.

\section{Conclusions}
In summary, we computed $A^{b\bar{b}}_{FB}$ in the SM and for several
NP scenarios, carefully accounting for the $Z$-pole, which is in the
signal region for the $b\bar{b}$ asymmetry.  The largest SM
contribution to $A^{b\bar{b}}_{FB}$ near the $Z$-pole comes from tree
level exchanges of $Z$ and $\gamma^{\star}$.  While at higher
invariant mass, NLO QCD dominates the SM asymmetry.  Light NP,
$M_{\text{NP}} \lsim 150$ GeV, is needed to generate a $b\bar{b}$
asymmetry, which CDF would be able to distinguish from the SM.
$A^{b\bar{b}}_{FB}$ can be used to distinguish between competing NP
explanations of $A^{t\bar{t}}_{FB}$ based on how the NP interferes
with $s$-channel gluon and $Z$ exchange.

\begin{acknowledgments}
We thank Ezequiel \'{A}lvarez, Dante Amidei, Aneesh Manohar, Manuel Perez-Victoria, Jesse Thaler, Michael Trott, and Thomas Wright for helpful discussions.  This work has been supported in part by the U.S. Department of Energy under Grant No. DE-SC0009919.
\end{acknowledgments}

\bibliography{BBbar}

%
%
\end{document}